\documentclass[aps, prd, 10pt, twocolumn, longbibliography, amsfonts, amssymb, amsmath, showkeys, nofootinbib, superscriptaddress, floatfix]{revtex4-2}
\usepackage[english]{babel}
\usepackage[utf8]{inputenc}
\usepackage{amsthm}
\usepackage{mathtools,amsmath,bm}
\usepackage{physics}
\usepackage{xcolor}
\usepackage{graphicx}
\usepackage{adjustbox}
\usepackage{placeins}
\usepackage[T1]{fontenc}
\usepackage{lipsum}
\usepackage{csquotes}

\usepackage{placeins}
\usepackage{tabularx}
\usepackage{hhline}
\usepackage{xcolor}
\usepackage[breaklinks, plainpages=false, colorlinks=true, anchorcolor=cyan, linkcolor=red, citecolor=blue, urlcolor=blue, bookmarks=false]{hyperref}
\usepackage{afterpage} 
\usepackage{array, booktabs, makecell}
\usepackage{multirow, bigdelim}
\usepackage{colortbl}
\usepackage{xspace}
\usepackage[normalem]{ulem}

\usepackage{comment}

\def\Msolar{M\ensuremath{_\odot}\xspace}
\newcommand{\asharp}{A$^\sharp$\xspace}
\newcommand{\aplus}{A+\xspace}
 
\newcommand{\ecc}{ECC\xspace}
\newcommand{\5}{O5\xspace}
\usepackage[acronym]{glossaries}
\newcolumntype{P}[1]{>{\centering\arraybackslash}p{#1}}

\newacronym{ce}{CE}{Cosmic Explorer}
\newacronym{et}{ET}{Einstein Telescope}
\newacronym{cehs}{CEHS}{Cosmic Explorer Horizon Study}
\newacronym{nsf}{NSF}{National Science Foundation}
\newacronym{xg}{XG}{Next Generation}

\begin{document}
\include{aas_macros}
\title{Cosmic Calipers: Precise and Accurate Neutron Star Radius Measurements with Next-Generation Gravitational Wave Detectors}
\author{Sanika Khadkikar}
\affiliation{Institute for Gravitation and the Cosmos, Department of Physics, Pennsylvania State University, University Park, PA 16802, USA}

\author{Ish Gupta}
\affiliation{Institute for Gravitation and the Cosmos, Department of Physics, Pennsylvania State University, University Park, PA 16802, USA}
\author{Rahul Kashyap}
\affiliation{Institute for Gravitation and the Cosmos, Department of Physics, Pennsylvania State University, University Park, PA 16802, USA}
\affiliation{Department of Physics, Indian Institute of Technology Bombay, Mumbai 400076, India}
\author{Koustav Chandra}
\affiliation{Institute for Gravitation and the Cosmos, Department of Physics, Pennsylvania State University, University Park, PA 16802, USA}
\author{Rossella Gamba}
\affiliation{Institute for Gravitation and the Cosmos, Department of Physics, Pennsylvania State University, University Park, PA 16802, USA}
\affiliation{Department of Physics, University of California, Berkeley, CA 94720, USA}
\author{Bangalore S. Sathyaprakash}
\affiliation{Institute for Gravitation and the Cosmos, Department of Physics, Pennsylvania State University, University Park, PA 16802, USA}

\begin{abstract}
Gravitational waves from merging binary neutron stars carry characteristic information about their astrophysical properties, including masses and tidal deformabilities, that are needed to infer their radii.  In this study, we use Bayesian inference to quantify the precision with which radius can inferred with upgrades in the current gravitational wave detectors and next-generation observatories such as the Einstein Telescope and Cosmic Explorer. We assign evidences for a set of plausible equations of state, which are then used as weights to obtain radius posteriors. We find that prior choices and the loudness of observed signals limit the precision and accuracy of inferred radii by current detectors. In contrast, next-generation observatories can resolve the radius precisely and accurately, across most of the mass range to within $\lesssim 5\%$ for both soft and stiff equations of state. We also explore how the choice of the neutron star mass prior can influence the inferred masses and potentially affect radii measurements, finding that choosing an astrophysically motivated prior does not notably impact an individual neutron star's radius measurements.
\end{abstract}
\maketitle

%\tableofcontents
%%%

%%%
\section{Introduction}
\label{sec1}

\par
Neutron stars (NSs) offer a unique environment to investigate matter at densities that exceed those of atomic nuclei, providing crucial insights into how matter behaves under such extreme conditions \cite{Lattimer:2012nd, Lattimer:2015nhk, Ozel:2016oaf, Watts:2016uzu, Baym:2017whm, Oertel:2016bki, Paschalidis:2016vmz}. The NSs' equation of state (EoS) connects the microscopic properties of dense matter to macroscopic observables such as radii, masses, and tidal deformabilities. Astronomical observations by instruments like NICER \cite{2016SPIE.9905E..1HG} provide direct constraints on masses and radii, enabling the inverse problem of mapping these quantities to the EoS from the mass-radius ($m$-$R$) plane \cite{Miller:2019cac, Miller:2021qha, Raaijmakers:2019dks, Raaijmakers:2019qny, Raaijmakers:2021uju, Bogdanov:2019ixe, Bogdanov:2019qjb}. Similarly, terrestrial experiments, such as PREX, probe the neutron skin thickness of nuclei, which correlates with the NS radius \cite{Roca-Maza:2011qcr, Fattoyev:2017jql, Reed:2021nqk}.

In the same way, gravitational waves (GWs) from binary neutron star (BNS) mergers have added a complementary approach to constrain the NS radius. The observation of BNS merger events GW170817 \cite{LIGOScientific:2017vwq} and GW190425 \cite{LIGOScientific:2020aai} by the Advanced LIGO and Advanced Virgo observatories \cite{LIGOScientific:2014pky, VIRGO:2014yos} pioneered the inference of NS radii using GW signals. Observing electromagnetic counterparts, such as gamma-ray bursts and kilonovae, can further inform source parameter estimates when combined with GW data. For example, the gamma-ray burst GRB170817A and the kilonova AT2017gfo associated with GW170817 imposed tighter constraints on the EoS and consequently, $R$ \cite{LIGOScientific:2017zic, Radice:2017lry, Coughlin:2018fis, Raithel:2019uzi, Capano:2019eae}. For instance, recently \citet{Breschi:2024qlc} and \citet{Koehn:2024set} showed that combining GW and multi-messenger observations reduces uncertainty in $R$ to less than 10\% at the 90\% credible interval \cite{Raaijmakers:2019dks, Raaijmakers:2021uju, Ayriyan:2021prr, Essick:2020flb, Hu:2020ujf, Raithel:2019uzi, Miller:2021qha}. 

However, constraints from GW observations remain broad due to GW170817’s relatively low signal-to-noise ratio (SNR), a limitation primarily driven by the sensitivity of the detector network at the time. Planned upgrades to current GW detectors and upcoming new observatories aim to overcome these limitations. During the fifth observational run (\5), the LIGO-Virgo-KAGRA (LVK) \cite{LIGOScientific:2014pky, VIRGO:2014yos, KAGRA:2020tym} detectors might achieve \aplus sensitivity \cite{T1800042}. The planned \asharp upgrade will maximize the scientific potential of existing LIGO facilities \cite{T2200287} to their limits. Further improvements in sensitivity and bandwidth will be possible with next-generation (XG) detectors like the Cosmic Explorers (CEs) \cite{Reitze:2019iox, Evans:2021gyd, LIGOScientific:2016wof} and the Einstein Telescope (ET) \cite{Punturo:2010zza, Hild:2010id, ET:2019dnz}. These will not only lead to relatively higher BNS detection rates but also more precise parameter measurements as compared to current results \cite{Gupta:2023lga}.

Several studies~\citep{Tsui:2005zf, Agathos:2015uaa, Lackey:2014fwa, HernandezVivanco:2019vvk, Landry:2018prl, Chatziioannou:2020msi, Wysocki:2020myz, Golomb:2021tll, Landry:2018prl, Landry:2020vaw, Essick:2019ldf, Ray:2023upk} have employed hierarchical inference and Bayesian model selection to constrain the EoS and infer NS radii. Some studies have specifically examined the radius measurement capabilities of XG detectors using these methods. For example, ~\citet{Ghosh:2022muc} and ~\citet{Pradhan:2023zor, Pradhan:2023xtq} estimated the uncertainty on $R_{1.4}$\footnote{The radius of a 1.4 \Msolar NS} using 50 (10) simulated BNS events, finding uncertainties of $\sim 2-3\%$ when a single XG detector is used. Similarly, using methods presented in \citet{Finstad:2022oni}, \citet{Bandopadhyay:2024zrr} showed that it is possible to constrain the $R_{1.4}$ to less than 10 m when all the BNS events observed by the 40 km CE over a decade are considered.

Many of these studies report uncertainties for $R_{1.4}$, assuming that the radius remains nearly constant for NS masses between 1.1 and 1.6 \Msolar. This assumption simplifies the analysis but introduces an error of $\sim$ 200 meters, comparable to or larger than the uncertainties expected from XG detectors \cite{De:2018uhw}. While the choice of $1.4$ \Msolar as a reference mass is motivated by its prevalence in the observed NS populations and similar calculations can be performed for other reference masses, quoting radius uncertainties for a given NS mass does not provide a complete description. This is because the uncertainties in $R$ depend on the NS mass, and the commonly quoted uncertainty in $R_{1.4}$ may not apply to other masses. Therefore, evaluating the precision and accuracy of the NS radius measurement with XG detectors requires considering the mass dependence of these uncertainties.

\citet{Huxford:2023qne} for the first time constrained the uncertainty in the NS radius for the entire mass range allowed by the considered EoSs (APR4, APR3, and ALF2), where the lower mass limit is $1~M_\odot$ and the upper mass limit is determined by the respective EoS. The uncertainties were constrained to within $2\%$, $1\%$, and $0.5\%$ for configurations with one, two, and three XG observatories using Fisher methods. ~\citet{Walker:2024loo} also obtained an uncertainty of $\sim 200$\,m in the $1$--$1.97$ \Msolar range and $\sim 75$\,m in the $1.4$--$1.6$ \Msolar mass range by applying spectral decomposition~\cite{Lindblom:2010bb} to the SLy EoS and by using a Bayesian framework. 

This work presents an alternative methodology for constraining NS radii using an evidence-based Bayesian model selection framework with an XG network and two planned upgrades in the current GW detectors for reference. Model selection assumes that the EoS model with the highest evidence represents the actual EoS. However, this assumption is valid only if the selected set maps the EoS space continuously and includes an infinite number of models. A discrete set of EoSs could miss regions of the parameter space where viable EoSs may exist. To avoid this, we use the dense set of models from Ref.~\cite{LIGOScientific:2018cki} (publicly available at \cite{ParametrizedEoS_MaxMass}), obtained by re-analyzing GW170817 
using a spectral decomposition for the EoS. 
In doing so, we cover a physically motivated region of the $m-R$ space. 
%% In principle, any number or form of tabulated or parameterized EoSs can make up this set as long as it is dense enough. 
%% We choose this parameterization because it is rooted in constraints provided by a detected event, allowing this study to serve as a proof of principle. 
We perform model selection over $\sim$ 2300 EoSs and calculate their evidence. Radius samples are drawn based on these EoSs in proportion to their evidence, as described in Sec. \ref{sec4a}, and combined to construct a posterior on $R$.

Additionally, we study how NS mass priors affect radius inference for individual events. With only a few detected BNS mergers, many options exist for choosing NS mass priors. Using a prior that does not match the underlying population can bias the inferred mass values, which can further bias radius measurements. Previous studies \cite{Agathos:2015uaa, Wysocki:2020myz, Landry:2020vaw, Golomb:2021tll} have shown that the EoS or NS mass distribution will be biased if inferred independently of the other. This study does not aim to infer the EoS or the BNS population. Instead, we investigate whether changing the NS mass prior affects $R$ inference for individual events. While one might expect that using a prior unlike the underlying population would strongly influence $R$ measurements, we find that these changes do not significantly impact the resulting single event radius posteriors.

The rest of this paper is organized as follows. Sec. \ref{sec2} summarizes parameter inference methods used in this study. In Sec. \ref{sec3}, we elaborate on our simulation setup by describing the GW detector networks in Sec. \ref{sec3a}, the NS mass population used in Sec. \ref{sec3b}. and the prior choices in Sec. \ref{sec3c}. Sec. \ref{sec4a} presents our methodology for inferring NS radii using the mass and tidal parameters extracted from GW signals of BNS mergers and presents the specifics of the EoSs used for model selection in Sec. \ref{sec4b}. We analyze the inferred radius posterior distributions in Sec. \ref{sec5}, summarize our conclusions, and propose future research directions in Sec. \ref{sec:conclusions}.

%%%%%%%%%%%%%%%%%%%%%%%%%%%%%%%%%%%%%
\section{Brief Review of Gravitational Wave Parameter Estimation}
\label{sec2}
\subsection{Binary Parameters and Waveform Models}
%%%%%%%%%%%%%%%%%%%%%%%%%%%%%%%%%%%%%

The calibrated detector data, \( d(t) \), for our case, can be modeled as a linear combination of detector noise, \( n(t) \), and a deterministic gravitational wave (GW) signal, \( h(t; \vec{\Theta}) \) where the latter is parameterized by signal parameters \(\vec{\Theta}\). The signal itself is expressible as a sum of the two GW polarizations \(h_{+,\times}\),  
\begin{equation}  
    h(t; \vec{\Theta}) = F_+ h_+ + F_\times h_\times \, ,  
\end{equation}  
where \( F_{+/\times} \) are the detector's antenna response functions, which are dependent on the source's sky location (\(\alpha, \delta\)) and polarization angle (\(\psi\)).  

For quasi-circular binary neutron star (BNS) systems, the parameter set \(\vec{\Theta}\) comprises 13 parameters: six intrinsic (component masses \( m_i \), spin components aligned with orbital angular momentum \( \chi_i \), and tidal deformability parameters \( \Lambda_i \)) and seven extrinsic (luminosity distance \( D_L \), inclination \( \iota \), polarization angle \( \psi \), coalescence time \( t_c \), coalescence phase \( \phi_c \), and sky position \(\alpha, \delta\)).  

Waveform models translate these intrinsic parameters of a BNS system into GW signals using different approximations of Einstein's field equations. Post-Newtonian (PN) models, for example, employ weak-field, low-velocity expansions and are well suited for describing the motion of neutron stars and their corresponding GW emission in the adiabatic inspiral regime~\cite{Futamase:2007zz, Blanchet:2013haa, Porto:2016pyg, Levi:2018nxp}. Within this framework, relativistic corrections to the Newtonian solution are incorporated systematically order by order using the expansion parameter \( x= v^2/c^2 \), where \( v\) is the orbital velocity and \(c\) is the light's speed in the vacuum. This formalism can provide a fast-to-evaluate, closed-form expression of the GW signal in the frequency domain by splitting the signal into an amplitude $\tilde{A}(f)$ and an overall phase $\Psi(f)$, assuming the following schematic form:
\begin{equation}
    \tilde{h}(f) = \tilde{A}(f)e^{i \Psi(f)}~.
\end{equation}
To the leading order, \(\tilde{A}(f)\) depends on the chirp mass, \( \mathcal{M} = \eta^{3/5} (m_1+m_2) \), the symmetric mass ratio, \( \eta = m_1m_2/{(m_1+m_2)}^2 \), \( D_L \) and \(\iota\) whereas \( \Psi(f) \) has the following schematic form: 
\begin{equation}
\begin{split}
    \Psi(f) &= 2\pi t_c - 2\phi_c - \frac{\pi}{4} + \frac{3}{128\eta x^{5/2}} \\
    &\times \big(1 + \varphi_{\rm PP} +  \varphi_{\rm SO} + \varphi_{\rm SS} + \varphi_{\rm Tidal}\big).
\end{split}
\end{equation}
Here, \( \varphi_{\rm PP} \), \( \varphi_{\rm SO} \), \( \varphi_{\rm SS} \), and \( \varphi_{\rm Tidal} \) capture contributions from point-particle, spin-orbit, spin-spin, and tidal interactions, respectively. 

Tidal effects arise due to mutual deformations induced by the binary components' gravitational fields and are quantified by the tidal deformability parameter,  
\begin{equation}
\label{eq2}
    \Lambda_i \equiv \frac{\lambda_i}{m^5} = \frac{2}{3} k_2 \left ( \frac{c^2R_i}{Gm_i} \right )^5,
\end{equation}
where \(\lambda_i\) parametrizes how much the \(i^\mathrm{th}\) NS with mass $m_i$ and radius \(R_i\) deforms, \( k_2 \) is the second tidal Love number and $G$ is the gravitational constant \cite{Hinderer:2007mb, Flanagan:2007ix}. 

The effective tidal deformability depends on the mass ratio $q=m_2/m_1 \leq 1$ and characterizes the leading order tidal phase contribution,  
\begin{equation}
    \Tilde{\Lambda} = \frac{16}{13}\frac{(12q +1)\Lambda_1 + (12+q)q^4 \Lambda_2 }{(1+q)^5}.
\end{equation}
which appears as a 5PN correction compared to the leading Newtonian term and, therefore, becomes more important as the binary approaches the plunge phase, where velocities increase. Although tidal corrections appear at high post-Newtonian orders, their contribution can be measurable because NS's $\Lambda \gtrsim 100$ for realistic EoSs, contingent on the NS's mass~\citep{Vines:2011ud, Wade:2014vqa, Chatziioannou:2020msi}.
% By construction, \( \Tilde{\Lambda} \) is an intrinsic parameter.  

Alternative waveform modelling strategies provide more accurate descriptions. For example, the effective one-body formalism combines insights from the test-particle limit with PN results to construct a unified description of the two-body problem. The EOB method reformulates the system with an effective one-body Hamiltonian, wherein a test particle moves in a deformed metric \cite{Buonanno:1998gg, Buonanno:2000ef, Damour:2001tu}. Since EOB models are inherently time-domain and require solving coupled differential equations, they are computationally expensive.  

On the other hand, phenomenological models focus exclusively on describing the GW signal without explicitly solving equations of motion~\cite{Ajith:2009bn, Santamaria:2010yb, Pratten:2020fqn}. Constructed in the frequency domain, they are computationally efficient and, hence, widely used for data analysis. They model the inspiral waveform based on analytic PN information augmented with pseudo-PN terms, then calibrated to EOB and NR results. 

Our study here employs the \texttt{IMRPhenomPv2\_NRTidalv2} waveform model \cite{Dietrich:2018uni, Dietrich:2019kaq}, an extension of \texttt{IMRPhenomPv2} \cite{Khan:2015jqa, Husa:2015iqa}. This model incorporates tidal-phase and amplitude corrections, which are constructed using Padé resummation of PN terms (up to 2.5PN adiabatic and 3.5PN spin-quadrupole contributions \cite{Dietrich:2019kaq}) and further refined through numerical relativity calibrations. These enhancements allow for improved modeling of BNS waveforms while maintaining computational efficiency.

\begin{figure}[b!]
    \centering    
    \includegraphics[width=0.46\textwidth]{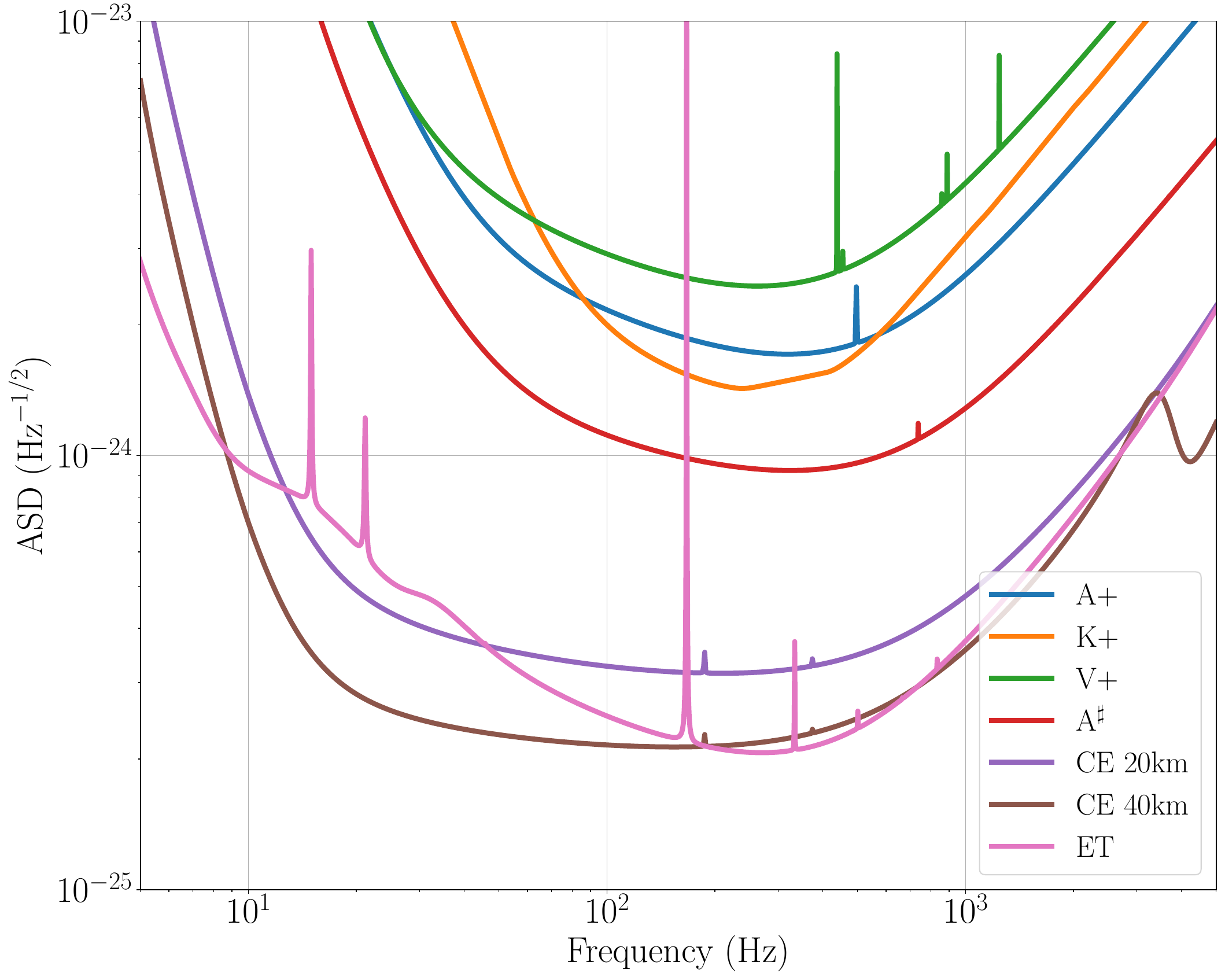}
\caption{Amplitude spectral density (ASD) of detector noise for LIGO (A+ and A\# sensitivity), Virgo (V+), KAGRA (K+), Cosmic Explorer (20 km and 40 km), and Einstein Telescope (ET) \cite{ET-0304B-22}.} 
\label{fig:asds}
\end{figure}

\begin{figure*}[!t!]
    \centering
    \includegraphics[width=0.8\textwidth]{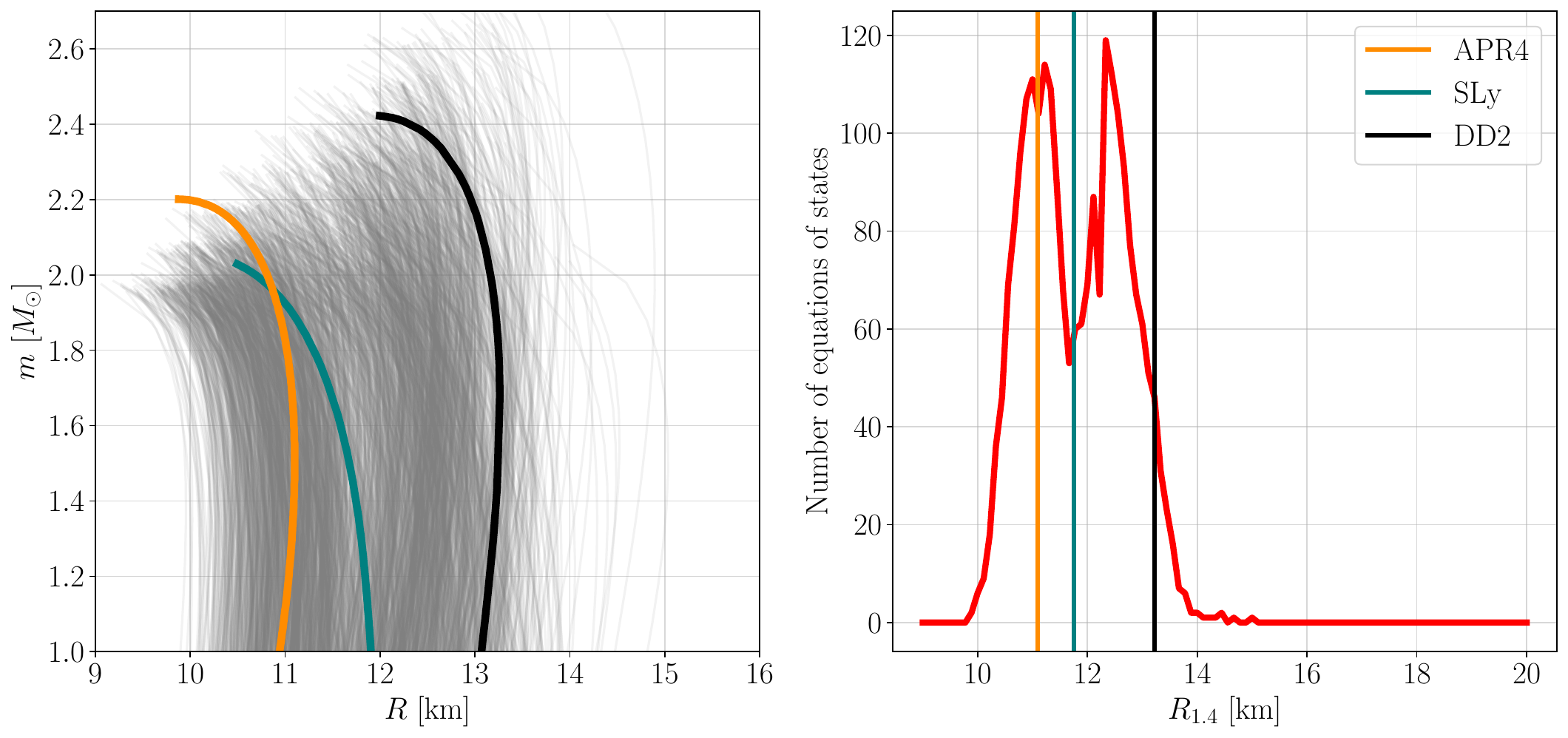}
\caption{\emph{Left panel}: $m\text{--}R$ curves for the EoSs used to calculate the evidence for EoS model selection. APR4 is shown in orange, SLy in teal, and DD2 in black. The gray EoS curves are the $\mathcal{O}(2300)$ EoSs parameterized using the spectral decomposition method for the GW170817 event as a part of the public data release for \cite{LIGOScientific:2018cki}. \emph{Right panel}: For the $\sim 2300$ EoSs considered, we calculate the number of EoSs within each radius bin for a $1.4\ M_{\odot}$ NS, providing insight into the density distribution of EoSs in the $m\text{--}R$ space. The APR4 EoS is represented in orange, SLy in teal, and DD2 in black. The density of EoSs is highest for radii larger than those predicted by APR4, while it is significantly lower for radii greater than those described by DD2.}
\label{fig:app3}
\end{figure*}

%%%%%%%%%%%%%%%%%%%%%%%%%%%%%%%%%%%%%%%%%%%%%%%%%%%%%%%%%%%%%%%
\subsection{Bayesian Inference}
%%%%%%%%%%%%%%%%%%%%%%%%%%%%%%%%%%%%%%%%%%%%%%%%%%%%%%%%%%%%%%%

We use Bayesian inference to estimate the binary parameters $\vec{\Theta}$ from the observed gravitational wave data. According to Bayes' theorem, the posterior probability distribution of the parameters $\vec{\Theta}$ given the data $d$ is:
\begin{equation}  
    p(\vec{\Theta} \mid d) = \frac{\mathcal{L}(d \mid \vec{\Theta}) \pi(\vec{\Theta})}{\mathcal{Z}_{GW}},  
\end{equation}  
where $\mathcal{L}(d \mid \vec{\Theta})$ is the likelihood function, $\pi(\vec{\Theta})$ represents the prior probability distribution for the parameters, and $\mathcal{Z}_{GW}$ is the model evidence given by:
\begin{equation}  
    \mathcal{Z}_{GW} = \int d\vec{\Theta}~\mathcal{L}(d \mid \vec{\Theta}) \pi(\vec{\Theta}).
\end{equation}  
The likelihood function quantifies how well a waveform $h(\vec{\Theta})$ describes the observed data and is typically modelled as a multivariate Gaussian. The choice of prior $\pi(\vec{\Theta})$ reflects our prior knowledge and constraints on the parameter space.

For our analysis, we use the Bayesian inference library \texttt{Bilby}~\cite{Romero-Shaw:2020owr, Ashton:2018jfp, lalsuite, swiglal} in conjunction with the nested sampling algorithm \texttt{dynesty}~\cite{Speagle:2019ivv} to obtain the posterior distribution of the model parameters. Given the high computational cost of analyzing long-duration BNS signals, we use the relative binning technique~\cite{Cornish:2010kf, Cornish:2021lje, Zackay:2018qdy} to accelerate likelihood evaluations. Relative binning improves computational efficiency as both the waveform evaluation and the sum over frequencies of the likelihood function are performed at a lower frequency resolution while maintaining the desired level of accuracy. We use the relative binning implementation in \texttt{Bilby} which is based on the functional form of the post-Newtonian expansion~\cite{Krishna:2023bug}.   

Further details regarding the specific configurations used for parameter estimation, including prior choices and sampling settings, are provided in Sec.~\ref{sec3c}.

%%%%%%%%%%%%%%%%%%%%%%%%%%%%%%%%%%%%%%%
\section{Simulation Setup}
\label{sec3}
%%%%%%%%%%%%%%%%%%%%%%%%%%%%%%%%%%%%%%%
\subsection{Network Configurations}
\label{sec3a}
%%%%%%%%%%%%%%%%%%%%%%%%%%%%%%%%%%%%%%%
We consider three different network configurations for our study as summarized in the following (see Fig. \ref{fig:asds} for their noise power spectrum):
\begin{itemize}
    \item 
    \textbf{\5}: We consider a network of LIGO Hanford, LIGO Livingston, Virgo, and KAGRA detectors operating at 
    \aplus sensitivity as the configuration for \5~\citep{KAGRA:2013rdx}.
    
    \item \textbf{\asharp}: The next upgrade of the LIGO detectors is the \asharp sensitivity \cite{T2200287}. For this configuration, in addition to LIGO Hanford and LIGO Livingston in the \asharp configuration, we have also considered LIGO India. LIGO India is a new GW interferometer currently being constructed in Aundha, India, and we assume it to be operating with \asharp sensitivity \cite{Unnikrishnan:2013qwa, Saleem:2021iwi, Pandey:2024mlo}. We do not include the Virgo or the Kagra detector for this network.
    
    \item \textbf{\ecc}: To analyze the most optimistic case for detecting radius uncertainties, we use XG observatories including one triangular ET \cite{Punturo:2010zza, Hild:2010id, ET:2019dnz},  one CE 20 km observatory, and one CE 40 km observatory \cite{Reitze:2019iox, Evans:2021gyd, LIGOScientific:2016wof}. We position the 40 km CE observatory in Hanford and the 20 km observatory in Livingston at the exact locations of the current LIGO interferometers. ET is positioned in Cascina in Italy, at the current location of the Virgo interferometer. 
    
\end{itemize}
%%%%%%%%%%%%%%%%%%%%%%%%%%%%%%%%%%%%%%%%
\subsection{Binary Population}
\label{sec3b}
%%%%%%%%%%%%%%%%%%%%%%%%%%%%%%%%%%%%%%%%

We conduct a simulation study by using sets of 100 synthetic BNS signals. This number is consistent with the cumulative number of detections expected within a redshift of $z=0.1$, assuming a BNS local merger rate density to be $320$ $ \rm Gpc^{-3} \rm yr^{-1}$\cite{LIGOScientific:2020kqk, Gupta:2023lga}. We sample masses for our simulated BNS signals from the galactic double Gaussian NS mass distribution model \cite{Antoniadis:2016hxz, Alsing:2017bbc, 2020RNAAS...4...65F, Shao:2020bzt, Landry:2021hvl} which has means $\mu_1 = 1.35 \, M_\odot$ and $\mu_2 = 1.8 \, M_\odot$, and standard deviations $\sigma_1 = 0.08 \, M_\odot$ and $\sigma_2 = 0.3 \, M_\odot$ consistent with the BNS mass population described in \cite{2020RNAAS...4...65F}. The NS masses range from 1 \Msolar up to the maximum mass allowed by the EoSs chosen to generate the simulated signals. The spin parameters are constrained to $\chi_i \in (-0.05, 0.05)$. The simulated binaries were distributed isotropically over the binary's orientation angle and sky location parameters.

For our simulated signals, we adopt three distinct equations of state (EoSs), strategically chosen to span the $m$-$R$ space supported by the GW170817 constraints \cite{LIGOScientific:2018cki, LIGOScientific:2019eut}. A \emph{soft} EoS, such as APR4, predicts relatively smaller tidal deformability ($\Lambda$) values for a given mass, corresponding to smaller neutron star radii. In contrast, a \emph{stiff} EoS, exemplified by DD2, yields larger $\Lambda$ values and, consequently, larger radii. The SLy EoS represents an intermediate scenario. The corresponding $m$-$R$ relationships for these EoSs are depicted in the left panel of Fig. \ref{fig:app3}. We generate simulated gravitational wave signals based on each EoS using the waveform model $\tt{IMRPhenomPv2\_NRTidalv2}$ and add them to data of each detector network described in Sec. \ref{sec3a}. We do not add Gaussian noise, ensuring that our analysis is performed in zero noise, and hence, our results represent the ensemble average over many Gaussian noise realizations~\citep{Nissanke:2009kt}. 

\begin{figure}[!t!]
    \centering    \includegraphics[width=0.46\textwidth]{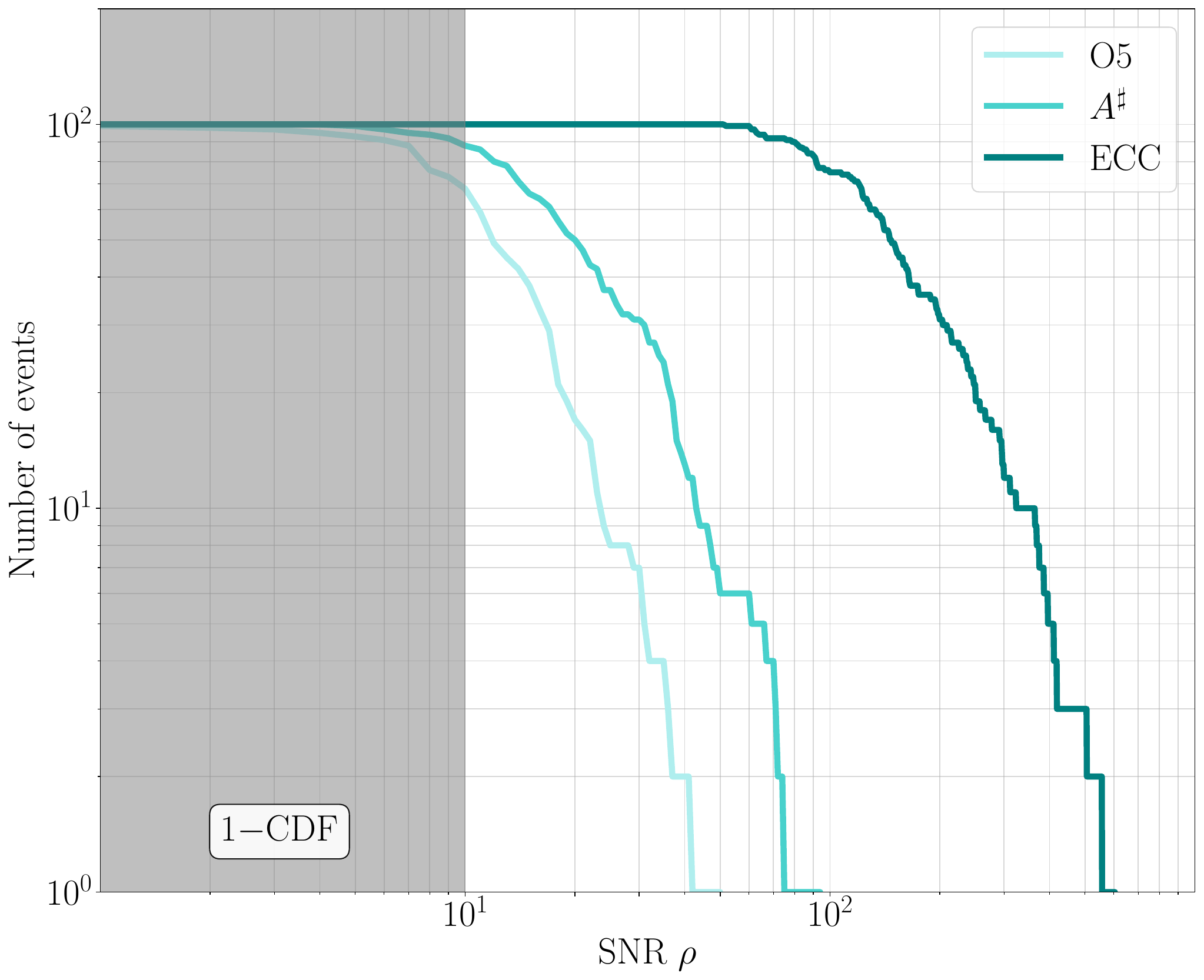}
    \caption{The scaled cumulative density function plot showing the trends in SNR $\rho$ of the 100 simulated events based on the SNR threshold of 10 for several GW detectors chosen. The plotted numbers of detected events can be obtained using any of the three EoSs considered; however, SLy is used here for demonstration purposes and is represented in teal. Different colors represent the different networks used for this study, namely \5, \asharp, and \ecc. While \ecc detects all of the simulated events, \asharp detects about 90\%, and \5 detects only about $70$\% of the simulated events. }
    \label{pop_dist}
\end{figure}

Fig. \ref{pop_dist} shows the number of detectable events as a function of the SNR. \ecc detects all events with $\mathrm{SNR} > 50$. In contrast, $A^\sharp$ and \5 networks miss approximately $10\%$ and $30\%$ of events respectively, assuming a network SNR threshold of 10.
%%%%%%%%%%%%%%%%%%%%%%%%%%%%%%%%%%
\subsection{Inference Settings and Prior Choices}
\label{sec3c}
%%%%%%%%%%%%%%%%%%%%%%%%%%%%%%%%%%
To obtain posterior samples, we utilize the relative binning implementation in \texttt{Bilby} with the \texttt{dynesty} sampler. Our analysis fixes the number of \texttt{walks} and \texttt{maxmcmc} at 100 and 5000, respectively. Additionally, we set the number of live points (\texttt{nlive}) to 1024 and the number of accepted steps per chain (\texttt{naccept}) to 60.  

Relative binning requires a tightly constrained prior on the chirp mass $\mathcal{M}$, as it assumes that the most probable waveforms used in the analysis do not differ significantly. Matched-filter searches provide a reasonably good estimate of $\mathcal{M}$, particularly for BNS signals, due to the high density of templates in this region of parameter space~\cite{Ewing:2023qqe}. Therefore, choosing a highly constrained prior on $\mathcal{M}$ is well justified. For the remaining parameters, we adopt the default prior choices for BNS analyses, with adjustments based on our simulated signals, as detailed in Table~\ref{tab:table1}~\citep{Veitch:2014wba, Romero-Shaw:2020owr, LIGOScientific:2020ibl}. We assume aligned spins, with maximum spin magnitudes within allowed regions of the employed waveform model. Additionally, we apply the \texttt{UniformSourceFrame} luminosity distance prior, which assumes a uniform prior in comoving volume $V_c$, scaled by a factor of $1/(1+z)$ to account for time dilation due to cosmic expansion, following the Planck 15 cosmology~\citep{Romero-Shaw:2020owr}.  

All prior choices are informed by theoretical models, physical constraints, and empirical data to reflect our prior knowledge or assumptions about the estimated parameters. When multiple viable priors exist, the Bayesian analysis pipeline can be rerun with different prior distributions. However, given the high dimensionality of the parameter space, this approach can be computationally expensive. Instead, posterior samples obtained using an initial prior $\pi_1(\vec{\theta})$ can be reweighted to reflect a different prior $\pi_2(\vec{\theta})$ using:  
\begin{equation}
    p_2(\vec{\theta}|d, H) = p_1(\vec{\theta}|d, H)\frac{\pi_2(\vec{\theta})}{\pi_1(\vec{\theta})}.
    \label{eq: reweighting}
\end{equation}  

This reweighting procedure is valid when the support of $\pi_2(\vec{\theta})$ is identical to or a subset of that of $\pi_1(\vec{\theta})$. In this study, we use prior reweighting to assess the impact of different mass priors on the inferred radius constraints for individual events. Specifically, we compare two priors: (i) a double-Gaussian prior matching the exact mass distribution used to generate our simulated BNS population and (ii) an uninformative prior that is uniform over the considered mass range. Since both priors share the same support, this approach allows for directly comparing their effects.

\begin{table}[ht!]
    \centering
    \begin{tabular}{|c|c|c|}
    \hline
         Parameter & Shape & Limits \\
    \hline
        $\chi_1,\chi_2$ & Uniform & (-0.1, 0.1) \\
        $cos(\theta_{JN})$ & Uniform & (-1, 1)  \\
        $\psi$ & Uniform & (0, $\pi$) \\
        $\alpha$ & Uniform & (0, $2\pi$)\\
        $\delta$ & Cosinusoidal & ($-\pi/2, \pi/2$)  \\
        $D_L$ & Uniform\ in Comoving\ Volume & (44, 475) Mpc \\
        $\Tilde{\Lambda}$ & Uniform & (0,5000) \\
        $\delta\Tilde{\lambda}$ & Uniform & $(-1000, 1000)$ \\
    \hline
    \end{tabular}
    \caption{Prior settings of signal parameters to study BNS signals.}
    \label{tab:table1}
\end{table}

%%%%%%%%%%%%%%%%%%%%%%%%%%%%%%%%%%%%%%%%%%%%%%%%%%
\section{Radius posterior and EoS model selection}
\label{sec4}
%%%%%%%%%%%%%%%%%%%%%%%%%%%%%%%%%%%%%%%%%%%%%%%%%%

We estimate posteriors for the parameter set $\vec{\Theta}$ by analyzing the simulated events using the parameter estimation methods outlined in Sec. \ref{sec2} and Sec. \ref{sec3}. However, Eq.~\eqref{eq2} indicates that the neutron star radius can only be inferred from these posteriors if the Love number $k_2$ is known. The value of $k_2$ depends on the underlying EoS governing the neutron star matter. Since the true EoS is unknown, we employ the Bayesian model selection pipeline \texttt{BEOMS} \cite{Kashyap:2024inprep} to compute the Bayesian evidence for each candidate EoS, denoted as $H_k$, among the $\sim 2300$ EoSs considered.

For this analysis, we work in a reduced parameter space $\vec{\theta} = {\tilde{\Lambda}, \mathcal{M}, \eta}$, chosen specifically to facilitate the computation of EoS evidence. We begin by obtaining $\tilde{\Lambda}_k$, which is determined from the posterior samples of the chirp mass $\mathcal{M}$ and the symmetric mass ratio $\eta$ under the assumption of the $k^{\rm th}$ EoS. This is the model against which we compare the posterior samples of $\tilde{\Lambda}$ obtained using parameter estimation.

Given that $\mathcal{M}$ is the most precisely estimated parameter in our inference, we simplify the analysis by fixing $\mathcal{M}$ to its mean value, $\overline{\mathcal{M}}$. This approximation reduces the dimensionality and computational complexity of the required integrals while maintaining sufficient accuracy for our model selection procedure.
%%%%%%%%%%%%%%%%%%%%%%%%%%%%%%%%%%%%%%%%%%%%%%%%%
\subsection{Evidence Calculation for EoSs}
\label{sec4a} 
%%%%%%%%%%%%%%%%%%%%%%%%%%%%%%%%%%%%%%%%%%%%%%%%%

We begin by marginalizing the likelihood $\mathcal{L}(d_i|\vec{\Theta})$ over the nuisance parameters to obtain $\mathcal{L}(d_i|\vec{\theta})$ in the reduced parameter space $\vec{\theta}$. For model selection, we use it to obtain the conditional likelihood $\mathcal{L}(d_i|\vec{\theta}, H_k)$:  

\begin{equation}
\mathcal{L}(d_i|\vec{\theta}, H_k) = \mathcal{L}(d_i|\vec{\theta}) \delta(\tilde{\Lambda} - \tilde{\Lambda}_k(\overline{\mathcal{M}}, \eta))~.
\end{equation}  

This ensures that only parameter space points supported by $H_k$ contribute where $H_k$ is the hypothesis that the $k^\mathrm{th}$ EoS describes the observed signal's tidal deformation parameters. Using Bayes' theorem, the posterior under $H_k$ becomes  

\begin{equation}
\label{posterior}
p(\vec{\theta} | d_i, H_k) = \frac{\mathcal{L}(d_i|\vec{\theta}) \delta(\tilde{\Lambda} - \tilde{\Lambda}_k(\overline{\mathcal{M}}, \eta))\pi(\vec{\theta}|H_k)}{\mathcal{Z}(d_i|H_k)},
\end{equation}  

where $\mathcal{Z}(d_i|H_k)$ is the model evidence,  

\begin{equation}
\label{evidence}
    \mathcal{Z}(d_i|H_k) = \int \mathcal{L}(d_i|\vec{\theta}) \delta(\tilde{\Lambda} - \tilde{\Lambda}_k(\overline{\mathcal{M}}, \eta))\pi(\vec{\theta}|H_k) d\vec{\theta}.
\end{equation}  

Rewriting the evidence in terms of the posterior,  
\begin{align}
    \mathcal{Z}(d_i|H_k) &= \int \mathcal{Z}_k \ \delta(\tilde{\Lambda} - \tilde{\Lambda}_k(\overline{\mathcal{M}}, \eta)) \, p(\overline{\mathcal{M}}, \eta, \tilde{\Lambda} | d_i) \, d\tilde{\Lambda} \, d\eta,
\end{align}  

where $p(\overline{\mathcal{M}}, \eta, \tilde{\Lambda} | d_i)$ is the joint posterior distribution, and  

\begin{equation}
    \mathcal{Z}_k = \int \mathcal{L}(d_i|\vec{\theta})\pi(\vec{\theta}) d\vec{\theta}
\end{equation}  

is computed analogously to $\mathcal{Z}_{GW}$ but in the reduced parameter space $\vec{\theta}$. We employ kernel density estimation (KDE) to obtain a smooth joint posterior $p(\overline{\mathcal{M}}, \eta, \tilde{\Lambda}|d_i)$, which, combined with the EoS, allows for evidence computation.  

To infer neutron star radii, mass posteriors $p(m | d_i, H_k)$ are mapped to radius posteriors $p(R | d_i, H_k)$ for each EoS. However, a physically meaningful radius distribution should incorporate contributions from all viable EoSs, weighted by their evidences. Instead of relying solely on the most probable EoS, we compute a model-weighted posterior  

\begin{equation}
    p(R|d_i) = \sum_k p(R|d_i, H_k) p(H_k|d_i),
\end{equation}  

where  

\begin{equation}
    p(H_k|d_i) = \frac{\mathcal{Z}(d_i|H_k)P(H_k)}{\sum_j \mathcal{Z}(d_i|H_j)P(H_j)}.
\end{equation}  

Assuming all EoS are equally likely, we get:  
\begin{equation}
    p(H_k|d_i) = \frac{\mathcal{Z}(d_i|H_k)}{\sum_j \mathcal{Z}(d_i|H_j)}.
\end{equation}  

We use $\mathcal{Z}(d_i|H_k)$ and compute the final model-weighted radius posterior as:
\begin{equation}
    p(R|d_i) = \sum_k p(R | d_i, H_k) \frac{\mathcal{Z}(d_i|H_k)}{\sum_j \mathcal{Z}(d_i|H_j)}.
    \label{eq18}
\end{equation}  

To implement Eq.~(\ref{eq18}), we perform KDE over each $p(R | d_i, H_k)$ and draw samples proportional to the corresponding EoS evidence within $R \in [5, 25]$ km. With $\sim2300$ EoSs considered, we generate $\sim$2.3 million samples to ensure sufficient representation. For each EoS, the sample count is  

\begin{equation}
    N_{k} = p(H_k|d_i)N_T,
\end{equation}  

where $N_T$ is the total sample count and $p(H_k|d_i)$ is the normalized EoS evidence. This process is repeated for all EoSs to construct the final radius posterior.  

Notably, EoSs support different mass ranges; for high-mass injections, not all posterior samples are compatible with every EoS. Our algorithm filters out unsupported mass samples from the radius inference, ensuring consistency. If an EoS has too few samples for KDE, it is excluded from inference, though this is rare given the extensive EoS set (see Fig.~\ref{fig:app3}). Evidences are renormalized after exclusion so that $\sum_k p(H_k|d_i) = 1$. Additionally, we examine the impact of mass prior choices on radius posteriors using the reweighting approach described in Sec.~\ref{sec2}.  

\begin{figure*}[!t]
    \centering
    \includegraphics[width=0.8\textwidth]{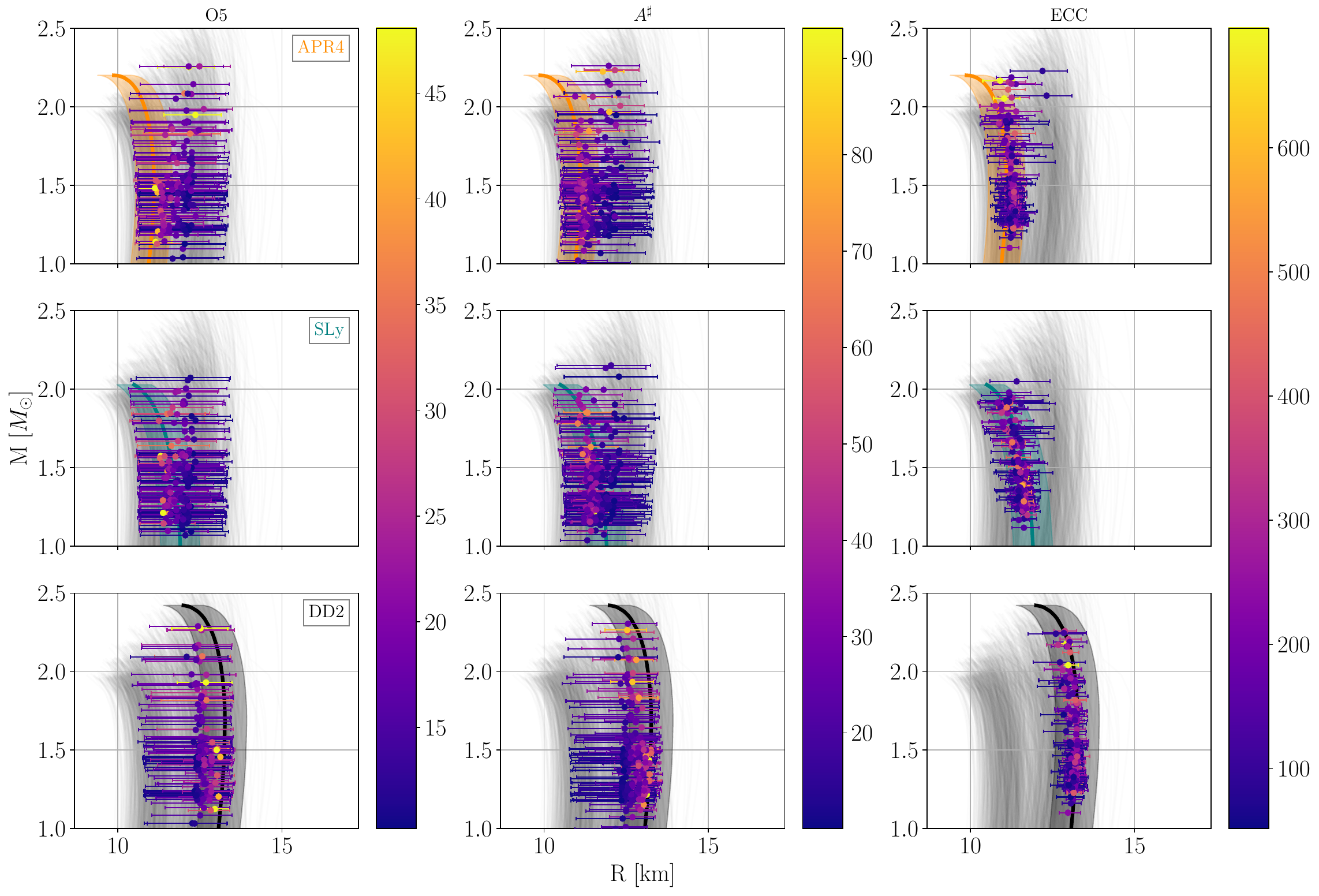}
    \caption{\textit{Top row}: Inferred median mass vs. radius posterior for the APR EoS (orange) for individual NSs in all detected signals. Radius uncertainties (90\% confidence intervals) are superposed on the EoS, with error bar colors representing SNR. Columns correspond to the \5 configuration (left), \asharp (middle), and \ecc (right). Gray curves depict EoSs from the spectral decomposition of GW170817. \textit{Middle and bottom rows}: Similar plots for the SLy (teal) and DD2 (black) EoSs. Biases in radius uncertainties stem from the non-uniform EoS set (Fig.~\ref{fig:app3}) and detector sensitivities. The \ecc configuration constrains $R$ most effectively, especially for the stiffer DD2 EoS, due to stronger tidal parameter effects.}
    \label{errors}
\end{figure*}

\par
\begin{figure*}[t!]
    \centering
    \includegraphics[width=0.8\textwidth]{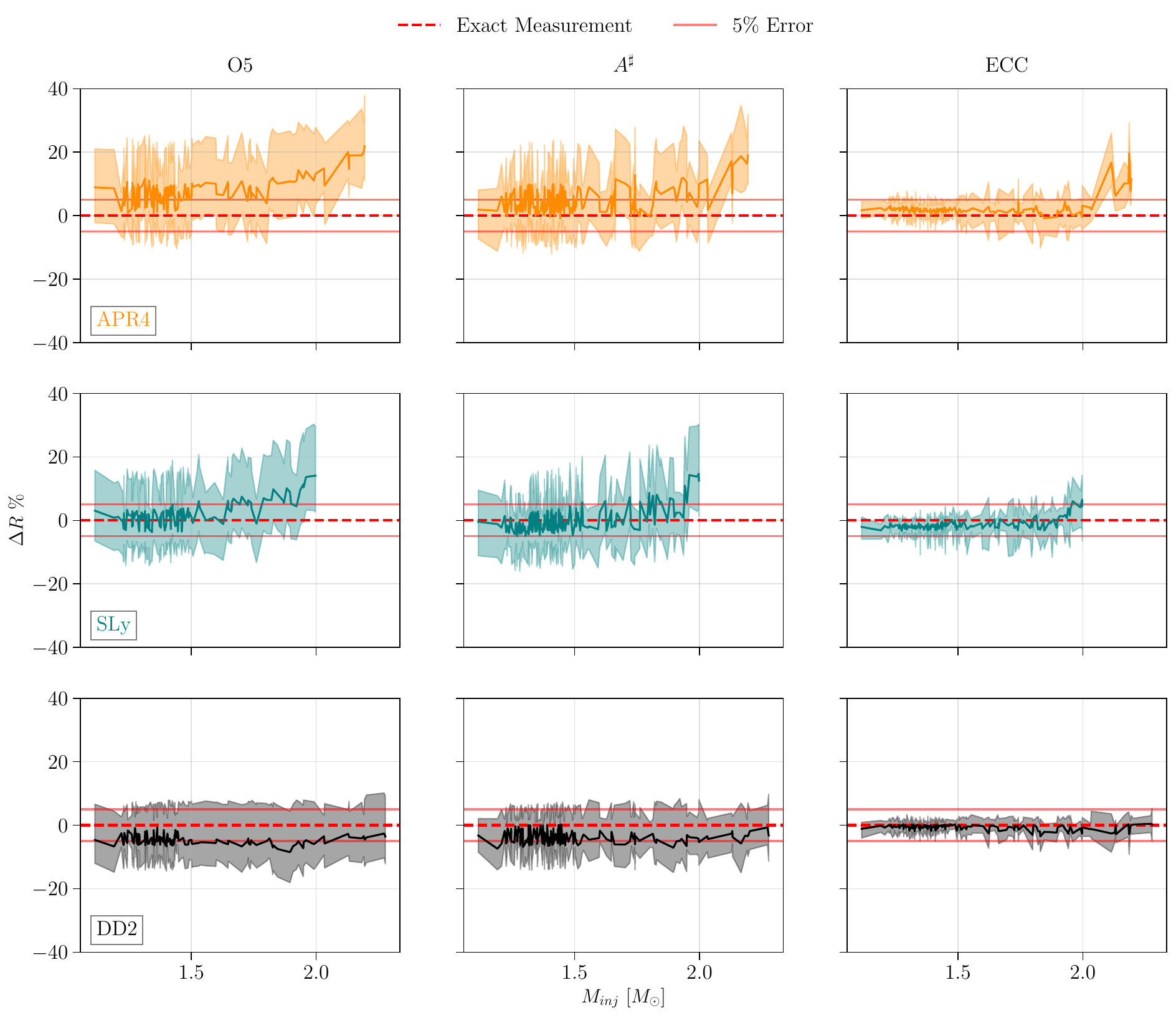}
    \caption{Fractional error percentage ($\Delta R$) in the inferred value of $R$ for the APR4, SLy, and DD2 (respectively top, middle, and bottom row) EoSs shown as a function of the injected NS masses. Different columns correspond to the considered detector sensitivities, increasing from left (\5) to right (\ecc). The solid colored line for every EoS shows $\Delta R$ for the median value of the inferred posterior and the band around it shows the 90\% credible interval. The red dashed line acts as a reference if there was no error and the red solid line shows 5\% errors in $\Delta R$. $\Delta R$ values corresponding to the \5 configuration are biased and the error is about 20\%. The results are similar for \asharp with the $R$ inference being inaccurate and imprecise. \ecc shows a remarkable increase in accuracy and precisions as $\Delta R$ is constrained very well across all EoSs to within 5\% for almost the entire mass range.} 
    \label{fig:bin_dist}
\end{figure*}

%%%%%%%%%%%%%%%%%%%%%%%%%%%%%%%%%%%%%%%
\subsection{EoSs used in simulations and model selection}
\label{sec4b}
%%%%%%%%%%%%%%%%%%%%%%%%%%%%%%%%%%%%%%%

We calculate model evidence for all simulated signals using $\sim$2300 additional EoSs beyond those used to create the simulated signals. These additional EoSs, shown as gray curves in Fig. \ref{fig:app3}, are constructed from GW170817 constraints using the spectral decomposition method as part of \citet{LIGOScientific:2017vwq, LIGOScientific:2018cki}.

Although the ensemble of EoSs effectively spans the $m$-$R$ space, including the regions around the three EoSs used for simulations, its density is uneven. To illustrate this, we bin the number of EoSs according to the radius they predict for a 1.4 \Msolar NS into $0.1 \ \rm km$ wide bins ranging from $9 \ \rm km$ to $20 \ \rm km$, as shown in the right panel of Fig.~\ref{fig:app3}. The density varies significantly, with more EoSs clustered around APR4 compared to SLy or DD2. This clustering increases the number of EoSs resembling APR4, which broadens the radius uncertainties for signals simulated using APR4. In contrast, the sparse representation of EoSs predicting radii higher than DD2 will cause a significant underestimation of $R$ for signals simulated using DD2. These non-uniformities in the EoS distribution introduce biases in the estimation of $R$. Addressing these biases will require either constructing a uniformly distributed EoS set or assigning a probability to each  EoS. We will explore these approaches in future studies.
%%%%%%%%%%%%%%%%%%%%%%%%%%%%%%%%%%%%%%%
\section{Results}
\label{sec5}
%%%%%%%%%%%%%%%%%%%%%%%%%%%%%%%%%%%%%%%

Building on the Bayesian methodology outlined in Sec. \ref{sec4}, we performed model selection across $\sim$2300 EoSs to compute their evidences for simulated BNS signals. These evidences were then used to derive radius posteriors, as detailed in this section.

%%%%%%%%%%%%%%%%%%%%%%%%%%%%%%%%%%%%%%%
\subsection{Radius uncertainties from upgraded LVK and XG detectors}
\label{sec5a}
%%%%%%%%%%%%%%%%%%%%%%%%%%%%%%%%%%%%%%%

We present the uncertainties in $R$, with error bars indicating the 90\% credible interval, for both companions from all BNS events with an SNR exceeding 10 in Fig.~\ref{errors}. These uncertainties are centered around the median value of $p(R|d_i)$. The SNR of each event is represented by a color bar (distinct for each detector network). A noticeable gap in the mass distribution around $2.1 M_\odot$ arises due to the double Gaussian population model, which provides little support in that region. To facilitate comparison, we overlay $5\%$ bands around the EoS, indicating the radius variations expected for a uniform $5\%$ change in the EoS.

Focusing on the \5 network, we observe significant uncertainties and systematic biases in most $R$ estimates. Given a fixed detector sensitivity, these uncertainties primarily stem from the EoS dependence of $R$ measurements. Softer EoSs, such as APR4 and SLy, predict smaller tidal deformability ($\Lambda$) values, resulting in a weaker tidal imprint on the gravitational wave (GW) signal and, consequently, broader uncertainties. This effect is particularly pronounced at higher masses, where tidal effects weaken as one approaches the binary black hole limit. Conversely, stiffer EoSs like DD2 predict larger $\Lambda$ values, leading to stronger tidal effects and tighter constraints on $R$. This trend is reflected in our results, with uncertainties being most significant for APR4, intermediate for SLy, and smallest for DD2, although the differences may not always be visually distinct. For example, the uncertainty in $R$ for the maximum-mass NS allowed by APR4 is approximately $16\%$, by SLy $14\%$, and by DD2 $11\%$.

Beyond uncertainties, systematic biases also affect $R$ estimates. For softer EoSs such as APR4 and SLy, the $R$ posteriors tend to be systematically overestimated. This arises from constraints on $\Lambda$: since $\Lambda$ must be positive, a hard boundary is imposed at $\Lambda = 0$. Weak constraints on $\Lambda$ and minimal tidal effects in the GW waveform—especially in less sensitive networks like \5—amplify this bias, causing overestimated $\Lambda$ values and, consequently, larger $R$ values.

However, this explanation alone does not fully account for the results, as we also observe a systematic underestimation of $R$ for stiffer EoSs like DD2. To better understand these biases, we refer to Fig.~\ref{fig:app3}, which depicts the distribution of EoS densities for a $1.4 M_\odot$ NS across different radii. The density of EoSs peaks around the APR4 model, followed closely by SLy and DD2. A larger proportion of EoSs resembling APR4 predict radii larger than those predicted by APR4 itself, leading to an overestimation of $R$. Conversely, fewer EoSs resemble DD2 and those that do predict smaller radii, resulting in an underestimation of $R$ for events analyzed with DD2.

These trends persist in the \asharp network, with only slight improvements in biases and uncertainties. This is primarily due to the relative sensitivity of the networks—\5 comprises four less sensitive networks, whereas \asharp consists of three more sensitive networks. Among these, the \ecc network stands out, as most $R$ measurements fall within the $5\%$ EoS bands, and its radius estimates exhibit reduced biases compared to other networks.

Figure~\ref{fig:bin_dist} summarizes our findings by illustrating the fractional percentage error $\Delta R$ as a function of injected mass across different EoSs and detector sensitivities. The key conclusion remains that for \5 and \asharp, the $R$ estimates suffer from imprecision and inaccuracy. In contrast, \ecc achieves superior accuracy, with $\Delta R$ remaining below $5\%$ for most masses. For context, the average values of $\Delta R$ for SLy are $10\%$, $9\%$, and $4\%$ for \5, \asharp, and \ecc, respectively. The increase in $\Delta R$ at higher masses, previously noted, is also evident in this plot. However, \ecc significantly mitigates this bias compared to other networks. While \5 and \asharp provide preliminary $R$ estimates, \ecc is required for reliable and precise constraints on $R$.

It is important to note that we do not specify a single numerical estimate for the lowest achievable radius uncertainty across networks, as uncertainties in $R$ are inherently mass-dependent. Figure~\ref{fig:bin_dist} demonstrates how uncertainties vary with mass, and quoting a single value would obscure this relationship. As the figure shows, for most EoSs, \ecc maintains uncertainties below $5\%$ across nearly the entire mass range. Thus, we recommend holistically interpreting Fig.~\ref{fig:bin_dist}, emphasizing the variation of uncertainties with mass rather than focusing on a specific value.

The accuracy of source mass estimates strongly depends on the precise inference of the luminosity distance and, consequently, the Hubble constant~\citep{Romero-Shaw:2020owr}. This dependence could introduce challenges in radius inference. However, since our analysis is limited to signals with redshift $z < 0.1$ and employs a consistent cosmological model for simulation and parameter estimation, these uncertainties do not significantly impact the inferred $R$ posteriors.

\subsection{Effect of mass priors}
\label{sec5b}
\begin{figure*}[t!]
    \centering    \includegraphics[width=0.8\textwidth]{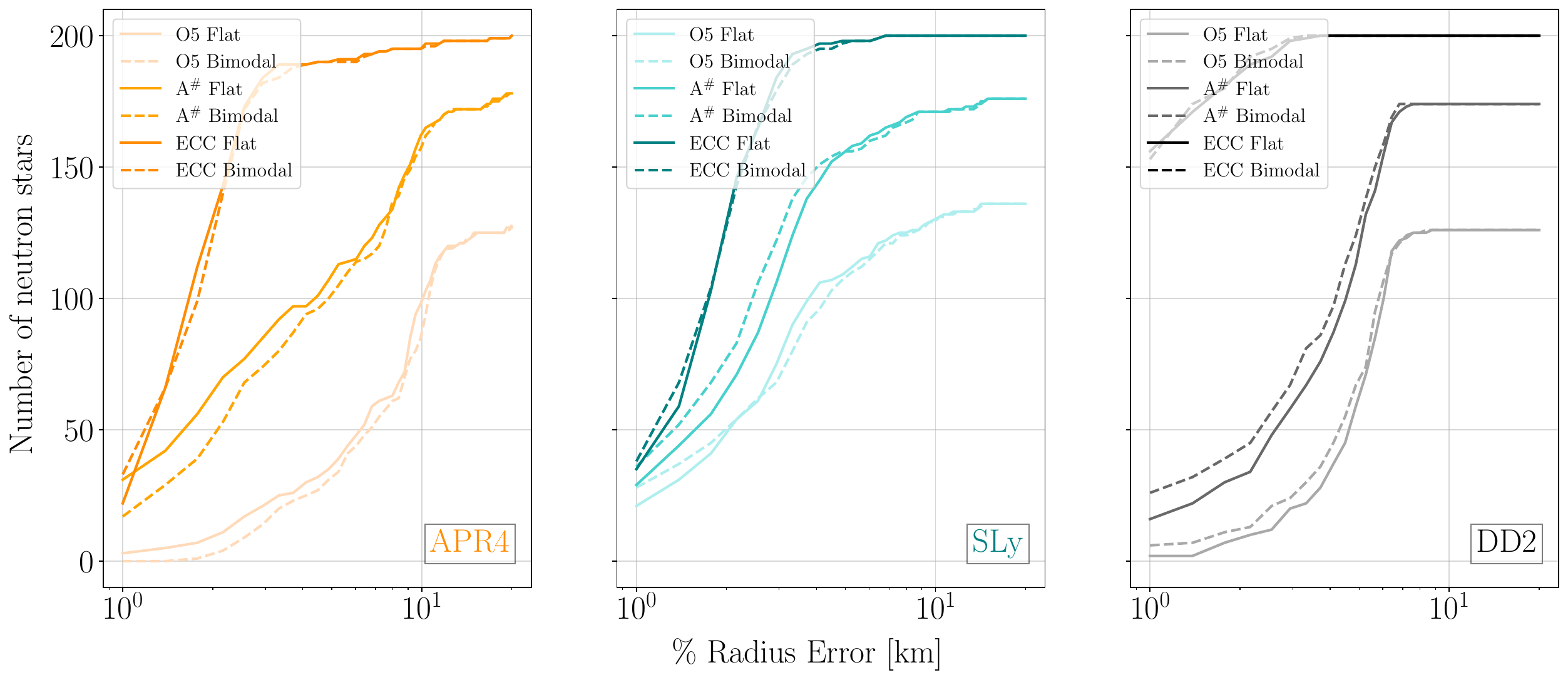}
    \caption{Comparing the radius uncertainties obtained with different priors on the NS mass for the APR4, SLy, and DD2 EoSs. Solid lines show the uncertainty for a flat prior, while dotted orange lines represent those obtained using a double Gaussian prior. Within a single panel, the color gradient indicates the detector sensitivity variation from \5 being the lightest to \ecc being the darkest. The obtained uncertainties don't change notably due to a change in the prior. It is also important to note that different networks detect different number of events, with \ecc detecting the most and \5 detecting the least, which is also reflected in these panels.} 
    \label{fig:priors}
\end{figure*}

Figure~\ref{fig:priors} presents the cumulative radius uncertainties for the two mass priors considered: a double Gaussian prior, closely resembling the true NS mass distribution, and a uniform mass prior. Intuitively, one might expect that a more realistic prior distribution would yield more precise radius measurements. However, as Fig.~\ref{fig:priors} illustrates, the uncertainties obtained using the two priors are nearly indistinguishable.

In the context of the \ecc network, where SNRs are exceptionally high (typically ranging from 100 to 700, Fig.~\ref{errors}), parameter estimates are tightly constrained, and the likelihood dominates the prior. As a result, the influence of the prior is significantly diminished, leading to a close agreement between posteriors derived from different priors. Since both priors impose similar mass ranges, and the component masses are well-measured, this agreement also extends to \5 and \asharp. We provide further details on this in Appendix~\ref{app1}.

%%%%%%%%%%%%%%%%%%%%%%%%%%%%%%%%%%%%%%%%%%%%%%%%%%%%%%%%%%%%%%%%%%%%%%%%%%%%%%%%%%%%%%%%%%%%%%%%%%%%%%%%
\section{Conclusions}
\label{sec:conclusions}
%%%%%%%%%%%%%%%%%%%%%%%%%%%%%%%%%%%%%%%%%%%%%%%%%%%%%%%%%%%%%%%%%%%%%%%%%%%%%%%%%%%%%%%%%%%%%%%%%%%%%%%%

GW observations provide a powerful means to study NS properties, particularly their radii. However, due to the EoS dependence of these measurements, direct inference of NS radii from BNS signals remains challenging. In this work, we employ a model selection framework using multiple EoS models derived from GW170817 to construct a model-weighted radius posterior, mitigating reliance on any single EoS assumption. This approach offers insights into the ultra-dense matter within NSs without requiring direct EoS inference.

Fig.~\ref{fig:bin_dist} demonstrates that the \ecc network provides precise NS radius measurements, with uncertainties ($\Delta R$) within ~5\% for stiffer EoSs like DD2. For softer EoSs such as APR4 and SLy, uncertainties remain within ~5\% in most cases. This underscores the capability of XG GW detector networks to refine NS property estimates. While the \5 and \asharp networks yield comparable uncertainties at times, their median radius values exhibit biases due to systematic overestimations of $\Lambda$ and the density of EoS models used in selection. Addressing these biases and improving uncertainty estimates for \5 and \asharp remains a key objective for future studies. Furthermore, combining GW-inferred radii with multi-messenger observations can enhance EoS constraints.

Additionally, we investigate the role of astrophysically informed priors in Bayesian inference. Contrary to expectations, our results indicate that such priors do not significantly improve the accuracy of radius estimates compared to flat priors. This suggests that radius precision is predominantly dictated by the signal's loudness and the analysis methodology rather than prior assumptions about NS masses.

We therefore believe that our study advances our ability to infer NS properties from GW observations, offering valuable implications for nuclear physics and astrophysics. The projected precision in radius measurements will be instrumental in future observational and theoretical efforts to constrain the EoS and understand NS interiors.

%%%%%%%%%%%%%%%%%%%%%%%%%%%%%%%%%%%%%%%%%%%%%%%%%%%%%%%%%%%%%%%%%%%%%%%%%%%%%
\section{Acknowledgments}
%%%%%%%%%%%%%%%%%%%%%%%%%%%%%%%%%%%%%%%%%%%%%%%%%%%%%%%%%%%%%%%%%%%%%%%%%%%%%
We thank Geraint Pratten, Justin Janquart, and Bikram Pradhan for reviewing the BEOMS pipeline, which was critical for the Bayesian model selection used in this study. We also thank Lami Suleiman for reviewing the manuscript and Jacob Goulomb, Isaac Legred, Katerina Chatziioannou, and Reed Essick for their insightful comments. This research is supported by the National Science Foundation's (NSF) LIGO Laboratory. SK, IG, and BSS are also supported by NSF awards PHY-2207638, AST-2307147, PHY-2308886, and PHY-2309064, as well as by grants OAC-2346596, OAC-2201445, OAC-2103662, OAC-2018299, and PHY-2110594, which provided access to the GWAVE cluster at PennState. RG acknowledges support from NSF Grant PHY-2020275 (Network for Neutrinos, Nuclear Astrophysics, and Symmetries (N3AS)). This document is also available on LIGO DCC with the document number P2400544. 

\bibliography{references}{}

\appendix
\section{Details about Mass Priors}

\label{app1}
\begin{figure}[!b]
    \centering
\includegraphics[width=\linewidth]{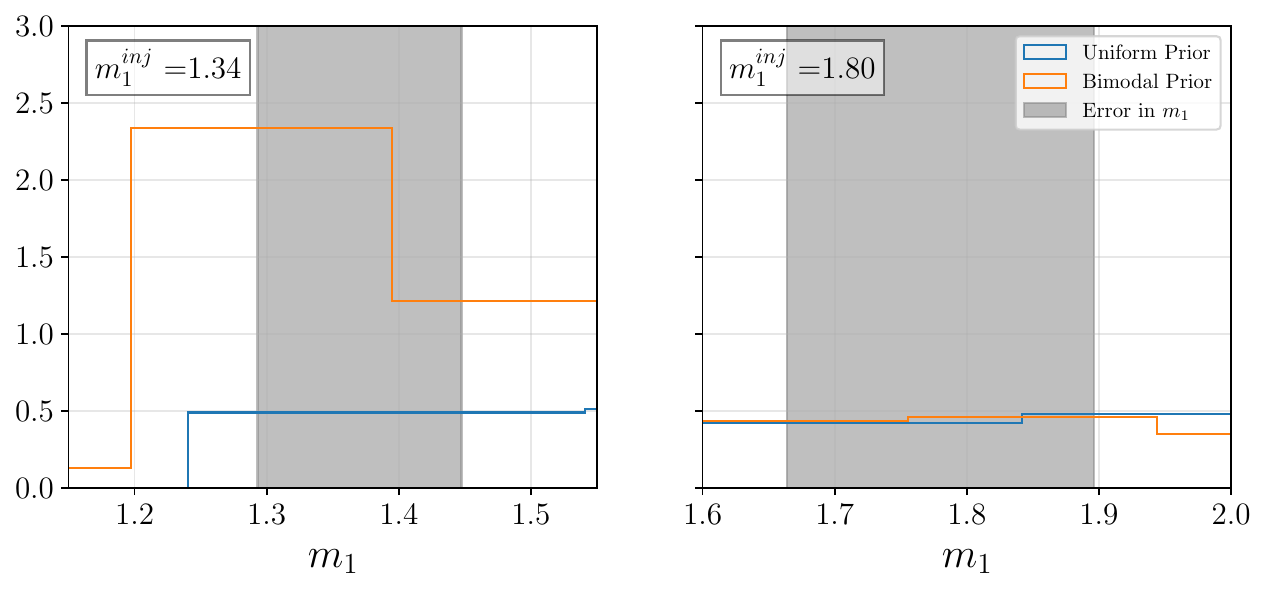}
    \caption{\emph{Left panel}: Prior samples drawn from the bimodal and uniform prior distributions for the heavier component mass $m_1$ for an example event having $m_1 = 1.34$. The gray band represents the uncertainty in the inferred value of $m_1$ for this event. \emph{Right panel}: A similar plot with another example event having higher $m_1 = 1.80$. The prior do not differ significantly from each other in the 1$\sigma$ range of the inferred posterior leading to little to no effects on the inferred radius uncertainty values.}
    \label{fig:appendix1}
\end{figure}
\begin{figure*}[!t!]
    \centering
    \includegraphics[width=0.8\linewidth]{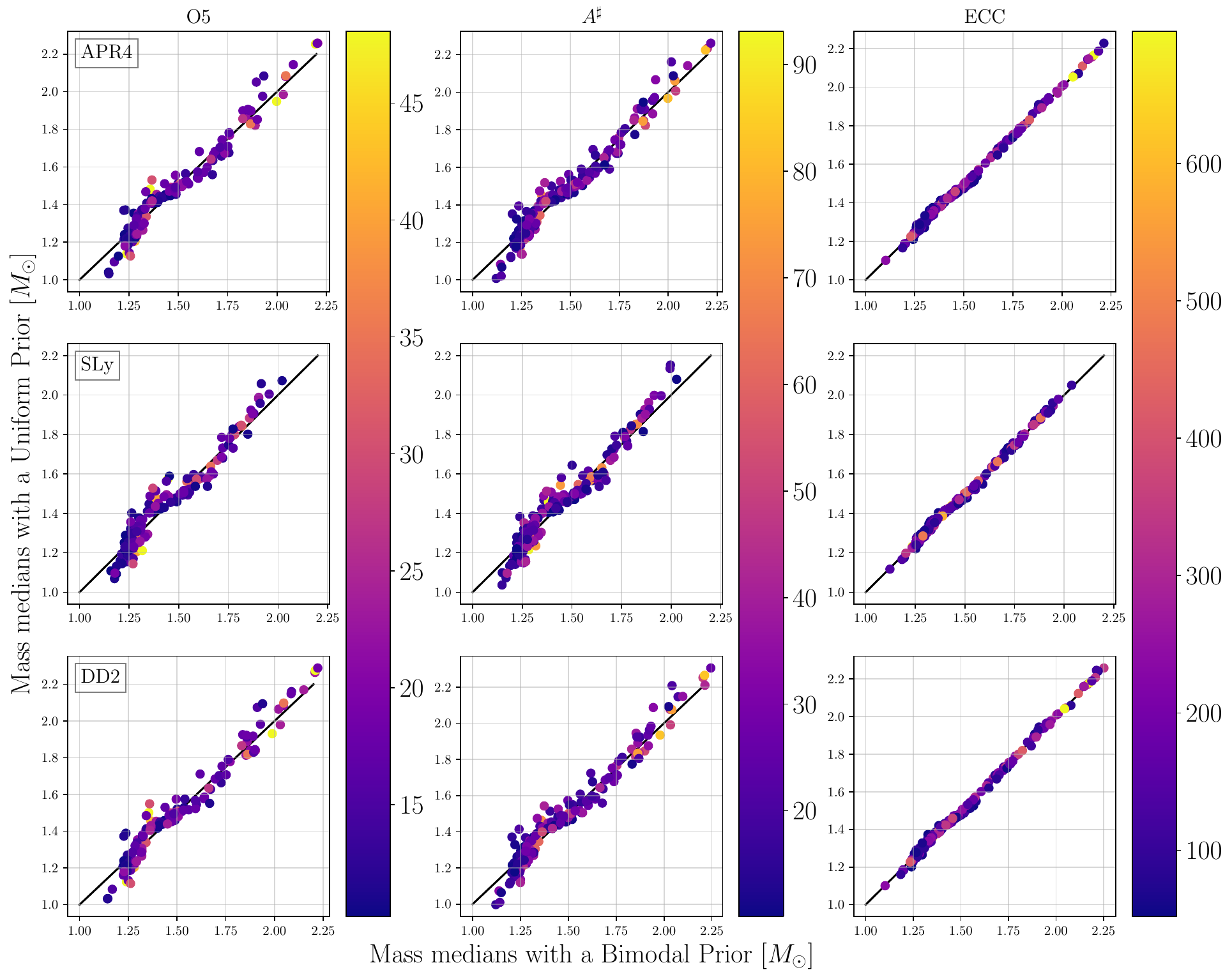}
    \caption{Comparison of the median values of the posterior distributions for the heavier component mass, $m_1$, obtained using the uniform and bimodal priors. Since the prior distributions do not vary significantly over narrow mass ranges, the median values from both distributions remain largely consistent. The color bar represents the SNR of the events, providing context for those that deviate from the equality line.}
    \label{fig:appendix2}
\end{figure*}

In Fig. \ref{fig:priors}, we compare the radius uncertainties obtained using the bimodal prior and the uniform priors. We conclude that on an individual event level, the uncertainties do not differ much for different detector sensitivities and EoSs. In this appendix, we present a few more nuances about the reasoning behind such trends.

To gain insight into the prior distributions, we analyze two events using the \5 sensitivity and the SLy EoS. These events are chosen so that the heavier component mass, $m_1$, corresponds to the two peaks of the bimodal distribution, where the uniform and bimodal priors are expected to differ the most. Figure \ref{fig:appendix1} compares the prior samples drawn from the bimodal and uniform distributions for $m_1$ and shows the error in the measurement of $m_1$ for both events. For the event at the secondary peak, with $m_1 = 1.8M_\odot$, the priors align in the range relevant for posterior inference. Even for the event at the primary peak, with $m_1 = 1.34M_\odot$, where the prior distributions are expected to differ the most, the variation remains small to impact the mass posteriors. The posteriors are well constrained, as indicated by the gray band. To affect the results, the priors would need to be significantly different in a narrow mass range, which is not the case in Fig. \ref{fig:appendix1}. As a direct consequence, Fig. \ref{fig:appendix2} shows that the median values of the inferred mass posteriors remain consistent across both priors. Therefore, the radius uncertainties inferred using these priors do not show significant differences.

\end{document}